\documentclass[lettersize,journal]{IEEEtran}

\usepackage[colorlinks,urlcolor=black,linkcolor=blue,citecolor=blue]{hyperref}

\def\BibTeX{{\rm B\kern-.05em{\sc i\kern-.025em b}\kern-.08em
    T\kern-.1667em\lower.7ex\hbox{E}\kern-.125emX}}

\usepackage{rotating}
\usepackage{amssymb}
\usepackage{algorithm}
\usepackage{algpseudocode}
\usepackage{color}
\usepackage{epsf}
\usepackage{psfrag}
\usepackage{epsfig}
\usepackage{amsfonts}
\usepackage{textcomp}
\usepackage{comment}
\usepackage{footnote}
\usepackage{tablefootnote}
\usepackage[flushleft]{threeparttable}
\usepackage{paralist}
\usepackage{mdwlist}
\usepackage{amssymb}
\usepackage{amsmath}
\usepackage{url}
\usepackage{verbatim}
\usepackage{alltt}
\usepackage{balance}
\usepackage{multirow}
\usepackage{epstopdf}
\usepackage{lipsum}
\usepackage{float}
\usepackage{booktabs}
\usepackage{slashbox}
\usepackage[table]{xcolor}
\usepackage{array}
\usepackage{boldline}
\usepackage{caption,setspace}
\usepackage{mathtools}
\usepackage[table]{xcolor}
\usepackage{csquotes}

\usepackage{dsfont}

\usepackage{amsthm}
\newtheorem{definition}{Definition}

\usepackage{stackengine}[2013-10-15]

\usepackage[T1]{fontenc}
\usepackage{frcursive}
\usepackage{calligra}
\usepackage{wedn}
\usepackage{bm}
\usepackage{aurical}
\usepackage{upgreek}

\definecolor{darkgreen}{RGB}{0,204,0}

\usepackage{cite}
\usepackage{adjustbox}

\usepackage{esint}
\usepackage{tipa}
\usepackage{soul}
\usepackage{tikz}
\usepackage{listings}
\usepackage{multicol}
\usepackage{longtable} 
\usepackage{enumitem}

\newcolumntype{?}{!{\vrule width 1pt}}
\newcolumntype{+}{!{\vrule width 1.25pt}}

\def\hlineb#1{%
\noalign{\ifnum0=`}\fi\hrule \@height #1 %
\futurelet\reserved@a\@xhline}

\usepackage{cancel}

\usepackage{pgfplots}
\pgfplotsset{compat=1.18} 
\usepackage{mathrsfs}

\usetikzlibrary{arrows.meta, positioning, fit, shapes.multipart}
\usepackage{booktabs}  

\usepackage{ragged2e}  

\definecolor{darkgreen}{RGB}{0,153,0}
\definecolor{darkred}{RGB}{192,0,0}

\definecolor{orange1}{RGB}{224,112,34}

\usepackage{array}

\newcolumntype{J}[1]{>{\justifying\arraybackslash}p{#1}}

\begin{document}

\title{Agentic Multi-Persona Framework for Evidence-Aware Fake News Detection}

\author{Roopa Bukke,
Soumya Pandey, 
Suraj Kumar,
    Soumi Chattopadhyay,
    Chandranath Adak
\thanks{
Roopa Bukke, Soumya Pandey, Suraj Kumar, and Soumi Chattopadhyay are with the Dept. of CSE, Indian Institute of Technology Indore, Madhya Pradesh 453552, India. 

Chandranath Adak is with the Dept. of CSE, Indian Institute of Technology Patna, Bihar 801106, India.

Corresponding authors: S. Chattopadhyay (soumi@iiti.ac.in), C. Adak (chandranath@iitp.ac.in)
This work has been submitted to the IEEE for possible publication. Copyright may be transferred without notice, after which this version may no longer be accessible.
}
}


\maketitle

\begin{abstract}
The rapid proliferation of online misinformation threatens the stability of digital social systems and poses significant risks to public trust, policy, and safety, necessitating reliable automated fake news detection.  
Existing methods often struggle with multimodal content, domain generalization, and explainability. We propose AMPEND-LS, an agentic multi-persona evidence-grounded framework with LLM–SLM synergy for multimodal fake news detection. AMPEND-LS integrates textual, visual, and contextual signals through a structured reasoning pipeline powered by LLMs, augmented with reverse image search, knowledge graph paths, and persuasion strategy analysis. To improve reliability, we introduce a credibility fusion mechanism combining semantic similarity, domain trustworthiness, and temporal context, and a complementary SLM classifier to mitigate LLM uncertainty and hallucinations. Extensive experiments across three benchmark datasets demonstrate that AMPEND-LS consistently outperformed state-of-the-art baselines in accuracy, F1 score, and robustness. Qualitative case studies further highlight its transparent reasoning and resilience against evolving misinformation. This work advances the development of adaptive, explainable, and evidence-aware systems for safeguarding online information integrity.
\end{abstract}


\section{Introduction} 
Digital news ecosystems constitute a core component of modern computational social systems, influencing collective opinion formation, policy discourse, and societal stability. With the rapid expansion of online platforms, information now propagates across interconnected social networks at unprecedented speed and scale. While this connectivity enhances access to information, it simultaneously amplifies the spread of misinformation and fake news \cite{zhou2020survey}. Such distortions of public discourse can undermine institutional trust, polarize communities, and generate harmful societal consequences in domains including public health, politics, and safety. Consequently, robust and transparent fake news detection has become a critical problem for safeguarding the integrity and resilience of digital social systems \cite{shu2020fakenewsnet,das2025fakenewsdetectionllm}.
%
Early mitigation efforts relied on manual fact-checking, which, although reliable, cannot scale to the velocity and volume of online information flows \cite{graves2018understanding}.
Automated approaches reframed detection as supervised classification, leveraging linguistic cues, user behavior, and propagation patterns \cite{shu2017fake}. Deep learning further advanced these models by capturing complex textual, semantic, and contextual dependencies \cite{zhou2020survey}. Yet, the rapid evolution of misinformation and heterogeneous news sources reveals persistent gaps, particularly in multimodal integration, cross-domain generalization, and reliance on resource-intensive annotations.

Over time, methods have evolved from text-only classifiers to multimodal, domain-adaptive, and large language model (LLM)-driven paradigms. Multimodal approaches, from early joint encoders EANN \cite{wang2018eann}/ MVAE \cite{khattar2019mvae} to contrastive and graph-enhanced models \cite{wei2025structure,10.1145/3581783.3612423}, demonstrate the value of heterogeneous cues, but suffer from fixed encoders, shallow fusion, and costly pipelines. Domain adaptation methods \cite{WU2025111485,10.1145/3664647.3681317} mitigate distribution shifts yet depend on domain metadata and remain brittle under unseen domains. More recently, LLM-based methods \cite{hu2025synergizingllmsgloballabel,modzelewski2025pcotpersuasionaugmentedchainthought} enhance reasoning and semantic understanding, but face high computational costs, weak multimodal integration, and shallow, non-evidence-grounded explanations, limiting use in high-stakes settings. Collectively, existing approaches capture correlations but lack scalability, robustness, and evidence-grounding transparency \cite{harris2024review}.

To address these challenges, we propose an explainable, evidence-grounded multi-persona agentic framework for multimodal fake news detection. Our framework integrates evidence retrieval through text-based web search and reverse image search, with each piece of evidence evaluated by a composite reliability score combining lexical and semantic similarity, domain credibility, and temporal consistency. In parallel, a knowledge graph (KG) encodes entity–relationship structures to provide factual grounding. These heterogeneous signals are synthesized by a multi-persona agentic module that engages in multi-round, persona-driven questioning to iteratively refine its contextual memory. Building on this, we design a modular classification pipeline where an LLM-based classifier produces pseudo labels with justifications, and a lightweight small language model (SLM) distills both the agent’s reasoning and LLM outputs. Importantly, for uncertain cases, we add persuasion-based analysis that detects manipulative strategies, computes a persuasion index, and refines reasoning. This design ensures low latency, high efficiency, and strong explainability, advancing multimodal fake news detection. 
Our key \textbf{contributions} are summarized below:

\emph{\textit{(i) Complementary Agentic Reasoning with Persuasion Refinement and SLM Integration}}: We develop an LLM-based multi-persona reasoning framework where specialized agents (supervisor, journalist, legal, scientific) iteratively interrogate evidence. Uncertain cases are re-evaluated using a persuasion-knowledge module that detects manipulative strategies. The distilled reasoning is transferred to a lightweight SLM classifier, making the pipeline both interpretable and computationally efficient.

\emph{\textit{(ii) Cross-Modal Evidence Retrieval with Reliability Fusion}}: To support agentic reasoning, we design a robust evidence retrieval pipeline spanning text, images, and knowledge graphs. Candidate documents are ranked using a hybrid reliability score that fuses semantic similarity, domain credibility, and temporal consistency, ensuring complementary signals from multiple modalities reinforce factual grounding.

\emph{\textit{(iii) Unified Evaluation of Complementary Strategies}}: Experiments on benchmark datasets and real-world case studies, we show that the combination of agentic reasoning, persuasion refinement, and reliability-guided multimodal evidence provides consistent improvements over individual baselines. Our component-wise analysis further indicates that each strategy contributes in complementary ways toward enhancing accuracy, efficiency, and interoperability.


The rest of this paper is organized as follows. Section \ref{sec:related_work} reviews related work. Section \ref{sec:problem_formulation} introduces the problem formulation, and Section \ref{sec:method} details the proposed methodology. Section \ref{sec:result} presents the experimental results and analysis. Limitations and future research directions are discussed in Section \ref{limitation},  
followed by the conclusion in Section \ref{sec:conclusion}.

\section{Related Work} \label{sec:related_work}
Fake news detection is a challenging classification task requiring reasoning over diverse information sources. Early methods struggled with multimodal evidence, while recent baselines integrate text and images but rely on simple fusion. Current advances can be broadly grouped into three categories: multimodal techniques \cite{khattar2019mvae,wei2025structure}, domain adaptation methods, and advanced language model–based approaches.

\emph{(i) Multimodal Fake News Analysis:} 
Early multimodal methods, such as EANN \cite{wang2018eann} and MVAE \cite{khattar2019mvae}, combined textual and visual features but relied on fixed encoders with limited adaptability. Subsequent works tackled cross-modal inconsistencies via similarity- and ambiguity-aware fusion (SAFE \cite{zhou2020similarity}, CAFE \cite{chen2022cross}), while SpotFake \cite{singhal2019spotfake} integrated BERT for text yet retained shallow fusion. More recently, contrastive learning has dominated, with StruACL \cite{wei2025structure}, COOLANT \cite{10.1145/3581783.3613850}, and MFCL \cite{CHEN2025112800} improving alignment through adversarial or contrastive objectives, and GAMED \cite{10.1145/3701551.3703541} exploiting mixture-of-experts for feature discrimination. Graph-based approaches enhance this by incorporating semantic and knowledge graphs (HSEN \cite{10.1145/3581783.3612423}, KAMP \cite{10.1145/3696410.3714532}, GS2F \cite{10.1145/3708536}). Transformer-based architectures further refine fusion with hierarchical attention \cite{10.1145/3581783.3612423,10.1145/3581783.3613850,10.1145/3696410.3714532,li-etal-2025-imrrf,10834537}. Despite these advances, most methods still rely on pretrained encoders, incur high computational cost, and lack explicit evidence-grounded reasoning, explainability, and scalability, limiting robustness in high-stakes misinformation detection.
\begin{figure*}[!t]
    \centering
    \includegraphics[width=0.9\linewidth]{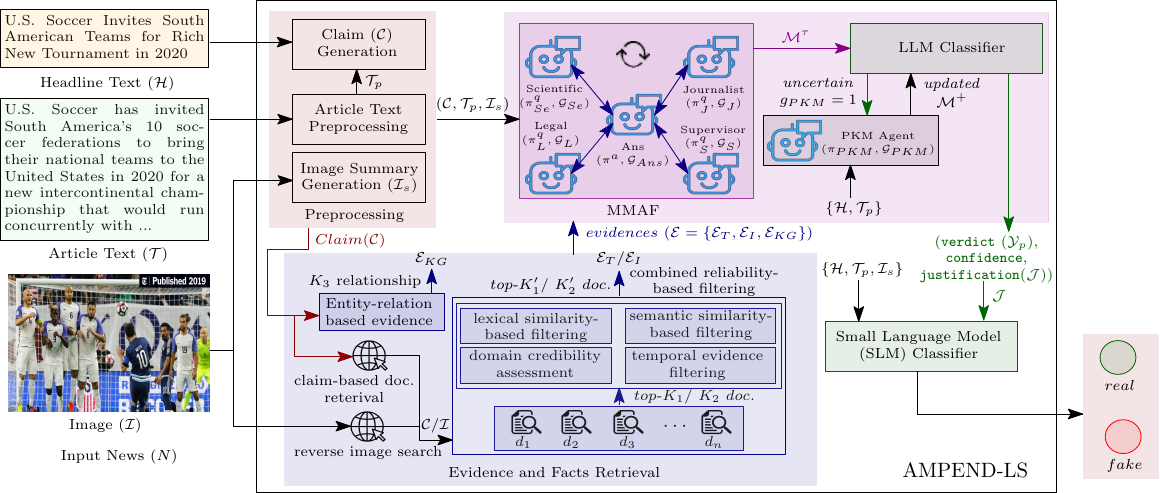}
    \caption{Proposed Framework: AMPEND-LS}
    \label{fig:arch}
\end{figure*}

\emph{(ii) Domain Adaptation-based Methods:} 
Domain adaptation is extensively studied to handle distribution shifts in fake news detection \cite{wei-etal-2025-cross}. Recent methods include hierarchical and transformer-based architectures such as HFGD \cite{WU2025111485}, MMDFND \cite{10.1145/3664647.3681317}, BREAK \cite{10834537}, and StruACL-TGN \cite{wei2025structure}, which leverage attention or structure-aware mechanisms for cross- and multi-domain settings. Transfer learning approaches like MMHT \cite{10.1145/3696410.3714517} disentangle veracity-relevant from domain-specific features, while PFND \cite{11027119} applies graph convolutional networks to combine content and fact-based evidence. Generative models such as MVAE \cite{khattar2019mvae} and GSFND \cite{tong2025generate} align multimodal features or augment training data. However, these methods often rely on domain labels or engagement signals, incur high cost, and lack interpretability, robustness, and evidence-grounded reasoning in mismatched or unseen domains.

\emph{(iii) Advance Language Model-based techniques:} 
LLMs have advanced fake news detection through reasoning-augmented and robustness-driven strategies. GLPN-LLM \cite{hu2025synergizingllmsgloballabel} incorporates global label reasoning, while PCoT \cite{modzelewski2025pcotpersuasionaugmentedchainthought} augments chain-of-thought prompting with persuasion cues. Robustness has been pursued via hallucination informed reasoning (IMRRF) \cite{li-etal-2025-imrrf}, adversarial style generation (AdStyle) \cite{10.1145/3696410.3714569}, and laundering-resistant detection \cite{das2025fakenewsdetectionllm}. Retrieval and collaboration approaches include self-reflective retrieval (SR3) \cite{Zhou_Zhang_Tan_Zhang_Li_2025} and multi-agent LLM–SLM co-learning \cite{10704605}. Data augmentation methods such as GSFND \cite{tong2025generate} and TripleFact \cite{xu2025triplefact} employ generative sampling and evidence-grounded triples. Yet, LLM-based systems remain limited by high computational cost, shallow multimodal integration, vulnerability to data contamination, and difficulties adapting to rapidly evolving misinformation.

\textit{\textbf{Positioning of our Work:}} 
To overcome the constraints of current approaches, we propose AMPEND-LS, an evidence-grounded, multi-persona agentic framework for robust fake news detection. Unlike domain adaptation methods that depend on domain labels or user engagement, and multimodal models limited by shallow fusion or costly encoders, AMPEND-LS explicitly integrates textual, visual, and contextual evidence with structured reasoning. A composite reliability score fuses lexical–semantic similarity, domain credibility, and temporal consistency, providing fine-grained veracity signals. Multi-persona agentic reasoning enables iterative claim validation through diverse expert roles, generating transparent justifications. This reasoning is distilled into a hybrid classification pipeline, where an LLM classifier captures rich contextual memory and a lightweight SLM provides efficient, uncertainty-aware verification. By combining evidence fusion, agentic reasoning, and scalable classification, AMPEND-LS achieves stronger factual grounding, interpretability, and adaptability than existing methods, positioning it as a deployable and trustworthy framework for combating misinformation in dynamic, multi-domain environments.

\section{Problem Formulation} \label{sec:problem_formulation}

Given a news article $N$, represented as $N = (\mathcal{H}, \mathcal{T}, \mathcal{I})$,
where $\mathcal{H}$ denotes the headline, $\mathcal{T}$ the article body text, and $\mathcal{I}$ an associated image.  
Each article has a ground-truth label
$
\mathcal{Y} \in \{REAL, FAKE\}.
$

 \emph{Challenges:} 
Detecting fake news is challenging due to:  
(i) the \emph{multimodal nature} of news, requiring joint reasoning across text and images,  
(ii) the presence of \emph{misleading or manipulated evidence}, which complicates credibility assessment, and  
(iii) the \emph{need for explainability}, where mere classification is insufficient without transparent reasoning.

\emph{Objective:} The goal is to learn a prediction function $f: N \mapsto \mathcal{Y}$, that is not only accurate but also \emph{evidence-aware} and \emph{explainable}.

\section{Proposed Methodology}
\label{sec:method}
Our proposed framework, AMPEND-LS (Fig.~\ref{fig:arch}), integrates textual, visual, and contextual cues into a unified pipeline designed to (i) handle multimodal inputs, (ii) ground predictions in credible evidence, and (iii) provide explainable reasoning.

The process begins with preprocessing, where headlines, article text, and associated images are refined into structured representations via claim extraction, text cleaning, and image summarization. An evidence retrieval and credibility assessment module then enriches the context with corroborating or refuting information from reliable sources, reducing reliance on potentially misleading inputs.
At the core lies the Multi-step Multi-persona Agentic Framework (MMAF), where LLM-powered agents adopt distinct personas (e.g., journalist, supervisor, legal analyst, scientific expert) and iteratively analyze claims through Q\&A interactions. This multi-perspective reasoning enables the capture of signals such as credibility, bias, and contextual consistency, producing both predictions and interpretable justification trails. Uncertain cases are not discarded but revisited with persuasion strategies to refine label quality and robustness.
Finally, to ensure efficiency, an SLM adaptation stage distills LLM-generated pseudo-labels and reasoning into a lightweight model. This preserves interpretability while enabling fast, resource-efficient inference with strong generalization.

\subsection{Preprocessing}\label{sec:preprocessing}
The stage processes heterogeneous news inputs into structured, semantically rich representations that enable effective evidence retrieval and reasoning.

\emph{(i) Article Text Preprocessing:}
Raw text $\mathcal{T}$ often contains artifacts (irregular punctuation, whitespace, noise). We apply standard cleaning operations to obtain a noise-free factual sequence $\mathcal{T}_p$, improving claim extraction and retrieval accuracy.

\emph{(ii) Claim Generation:}
Since headlines $\mathcal{H}$ are concise and $\mathcal{T}_p$ may contain extraneous details, we distill them into a concise, entity-aware claim $\mathcal{C} = f_1(\mathcal{H}, \mathcal{T}_p),$
where $f_1$ is an LLM-based claim generator that emphasizes salient entities and contextual cues. $\mathcal{C}$ facilitates precise evidence retrieval.

\emph{(iii) Image Summary Generation:} To capture visual signals, we extract image entities 
$\mathcal{V} = \{v_i\}_{i=1}^n$
(e.g., objects, locations, text) using Google Cloud Vision API (https://cloud.google.com/vision). These are embedded into an LLM prompt to generate summary $\mathcal{I}_s = f_2(\mathcal{I}, \mathcal{V}),$
where $f_2$ generates a coherent, human-readable description integrating $\mathcal{V}$ with salient visual content. $\mathcal{I}_s$ provides a structured textual representation of $\mathcal{I}$, suitable for fact-checking.

\noindent
The preprocessing stage yields $(\mathcal{T}_p, \mathcal{C}, \mathcal{I}_s)$, which together provide semantically rich, analyzable inputs for later stages.

\subsection{Evidence and Facts Retrieval}\label{sec:evidence_fact_module}
This module retrieves and validates external information to support or refute the generated claim $\mathcal{C}$, yielding structured evidence $\mathcal{E}$ for downstream agentic reasoning.

\emph{(i) Evidence Selection:}
To retrieve candidate evidence, we perform a web search on the generated claim $\mathcal{C}$ using the Google Cloud Custom Search API  (https://console.cloud.google.com), yielding a document set $\mathcal{D}=\{d_i\}_{i=1}^n$. To restrict the search space, we retain only the top-$K_1$ domains, denoted $top\text{-}K_1(\mathcal{D})$. 
For each candidate document, we compute two complementary similarity scores: 
\begin{equation}  \small 
s_1 = \sigma~ \bigl(BM25(\mathcal{C}, top\text{-}K_1(\mathcal{D}))\bigr)~;~ 
s_2 = BS(\mathcal{C}, top\text{-}K_1(\mathcal{D})); 
\end{equation} 
where, $s_1$ measures lexical relevance via BM25 \cite{bm25}, $s_2$ measures semantic similarity via BERT-Score \cite{bert_score}, and $\sigma$ is a normalization function. 
Formally, the evidence retrieval operator is defined as: 
{\small{$
f_{\text{retr}} : \mathcal{C}  \mapsto  \{(d_i, s_1(d_i), s_2(d_i))\}_{i=1}^{K_1}.
$}}

\noindent
This hybrid design balances lexical and semantic matching, ensuring that retrieved evidence is both textually aligned and contextually relevant for claim verification.

\emph{(ii) Domain Credibility Analysis:}
Textual similarity scores alone cannot ensure the trustworthiness of evidence, since unreliable or biased sources may exhibit high lexical or semantic overlap with the claim. To address this, we incorporate domain credibility as an additional signal of reliability.  

We adopt the Media Bias/Fact Check (MBFC) database (mediabiasfactcheck.com), which provides widely used independent assessments of media outlets by combining factual reporting quality with political/ideological bias ratings. MBFC is chosen for its comprehensive coverage and transparent methodology, making it a practical credibility reference.  

Each candidate domain is assigned a categorical credibility score DCS $\in \{\textit{High}, \textit{Medium}, \textit{Unknown}, \textit{Low}\}$, which we map to weights: 
\begin{equation} \scriptsize
s_3(d_i) = 
\{\text{High} \mapsto \alpha_h,
  \text{Medium} \mapsto \alpha_m,
  \text{Unknown} \mapsto \alpha_u, 
  \text{Low} \mapsto \alpha_l\}.
\end{equation} 
with ordering constraints: 
$\alpha_h {>} \alpha_m {>} \alpha_u {>} \alpha_l; ~ \alpha_j {\in} [0,1]; ~ j \in \{h,m,u,l\}$. 
Typical settings are $\alpha_h {\approx} 1$, $\alpha_l {\approx} 0$, and $\alpha_u {\approx} 0.5$ to avoid unfair exclusion of less popular domains.  
The credibility operator is therefore defined as: 
$f_{\text{cred}} : top\text{-}K_1(\mathcal{D}) \mapsto \mathbf{s}_3 = [s_3(d_1), \dots, s_3(d_{K_1})]$, which penalizes evidence from low-credibility sources while prioritizing reputable domains for claim verification.

\emph{(iii) Temporal Evidence Filtering:}
This step aims to retain evidence contemporaneous with the news event while filtering out outdated reports of similar past events. In the absence of a known publication date for the target article, we infer the temporal context of the event from the retrieved evidence itself. The key idea is that relevant sources should cluster within a dominant time window, whereas outdated reports of similar past events typically fall outside this cluster. To identify this window robustly, we use the median publication date $t_{\text{median}}$ of {the top-$K_1$ retrieved documents}, as it is less sensitive to outliers than the mean.  
Let $t_i$ denote the publication date of candidate document $d_i$ and $t_{\text{median}}$ the median date across top-$K_1$ retrieved documents. The temporal deviation is defined as:
$
\Delta_i = |t_i - t_{\text{median}}|.
$
The temporal consistency score is computed as:  
\begin{equation}\label{eq:date_consistency_inline}
\small
s_4(d_i) =
\begin{cases}
\max(\alpha_t, 1 - \Delta_i/365); & t_i \neq \text{unknown} \\
0; & \text{otherwise}
\end{cases}
\end{equation}

\noindent
where, scores close to $1$ indicate strong temporal alignment with the current event. A lower bound of $\alpha_t$, a hyperparameter, prevents discarding slightly delayed but relevant reports, while documents without valid timestamps are excluded. The temporal filtering operator is thus defined as
{\scriptsize{$
f_{\text{temp}} : top\text{-}{K_1}(\mathcal{D})  \mapsto  \mathbf{s}_4 = [s_4(d_1), s_4(d_2), \dots, s_4(d_{K_1})].
$}}

\noindent
The objective of this filtering is to prioritize evidence that co-occurs within the event’s active propagation window, thereby ensuring temporal relevance and excluding outdated or recycled content.

\emph{(iv) Reliability Score:}
For each $d_i \in top\text{-}K_1(\mathcal{D})$, the reliability score integrates all the above scores:
\begin{equation}
\scriptsize
\mathcal{R}(d_i) = \lambda_1 s_1(d_i) + \lambda_2 s_2(d_i) + \lambda_3 s_3(d_i) + \lambda_4 s_4(d_i); \quad \sum\nolimits_{j=1}^4 \lambda_j = 1
\end{equation} 
where $\lambda_j$ are tunable weights, for $j {\in} \{1, 2, 3, 4\}$.
Documents are ranked by their reliability score $\mathcal{R}(d_i)$, and the top-$K'_1$ are retained as credible evidence $\mathcal{E}_T$ for downstream misinformation analysis.

\emph{(v) Visual Evidence:}
Reverse image search is employed to trace the origin and context of images, enabling the retrieval of relevant external evidence to verify or refute potential fake news. We, therefore, perform reverse image search on the query image $\mathcal{I}$ using the Google Cloud Vision API. This yields a candidate set of documents with metadata: 
$\langle$\textit{image URL}, \textit{article URL}, \textit{web entities}, \textit{article title}, \textit{article summary}, \textit{date of publication}$\rangle$, 
from which irrelevant sources are discarded. 
For each candidate $d_i$, reliability is assessed using the same four scores as in textual evidence, lexical similarity, semantic similarity, domain credibility, and temporal consistency, resulting in an overall reliability score $\mathcal{R}(d_i)$.  

The top-$K_2$ candidates are ranked by $\mathcal{R}(d_i)$, and the top-$K'_2$ are retained as credible image-based evidence:
{\scriptsize{$
\mathcal{E}_I = \{d_i \mid d_i \in top\text{-}K'_2(\mathcal{D}_I)\}.
$}}

This ensures that only temporally relevant, domain-reliable, and contextually aligned visual sources contribute to downstream misinformation analysis.

\emph{(vi) Entity–Relation Evidence:}
To provide structured factual grounding for claim verification, we query external knowledge bases (Wikidata, DBpedia, Google Knowledge Graph) via LLMs to extract relational facts. These are represented as knowledge graph (KG) triplets
$
\langle s, r, o \rangle    \text{ with }    s,o \in \mathcal{E}_{ent},   r \in \mathcal{R}_{KG},
$
where $\mathcal{E}_{ent} = \{e_1, e_2, \dots, e_m\}$ denotes the set of entities extracted from the claim $\mathcal{C}$ (e.g., persons, organizations, locations, events), and $\mathcal{R}_{KG} = \{r_1, r_2, \dots, r_p\}$ denotes the set of factual relations supported by knowledge bases (e.g., \emph{foundedBy}, \emph{memberOf}, \emph{locatedIn}, \emph{dateOf}).  For example,
$\langle$ \text{NASA},   \textit{foundedBy},   \text{U.S. Government} $\rangle$. 
The retrieved triplets form $K_3$ number of entity–relation paths:
$
\mathcal{E}_{KG} = \{\langle s, r, o \rangle_k\}_{k=1}^{K_3}.
$

These structured relations are integrated with prior evidence sets to strengthen factual validation, enabling the framework to support or refute $\mathcal{C}$ against authoritative knowledge sources.

Collectively, this module constructs the structured evidence set
$
\mathcal{E} = \{\mathcal{E}_T, \mathcal{E}_I, \mathcal{E}_{KG}\},
$
This unified evidence representation $\mathcal{E}$ is forwarded to the next stage for advanced misinformation analysis.

\subsection{Multi-step Multi-persona Agentic Framework (MMAF)} 
\label{sec:multi_qa_framework} 
Building on the structured evidence $\mathcal{E}$, we introduce an MMAF for fake news detection.
MMAF organizes LLM agents into distinct roles: Supervisor, Journalist, Legal Analyst, and Scientific Expert, supported by an Answering Agent. Unlike single-agent prompting, these personas iteratively pose probing questions, with the Answering Agent delivering concise, evidence-grounded responses. A shared memory accumulates these interactions, enabling coordinated reasoning. This explicit role separation, disciplined answering, and iterative enrichment yield more robust, transparent, and evidence-aware verification than conventional approaches.

\begin{definition}[MMAF] \label{def:mp_framework}
An MMAF for evidence-aware fake news detection is a structured system of LLM-based agents. Formally, it is defined as:
$\mathcal{F} = \{A_{\text{Supervisor}}, A_{\text{Journalist}}, A_{\text{Legal}}, A_{\text{Scientific}}, A_{\text{Ans}}\}$, 
where each role-specific agent has the tuple structure
$A_i = (\mathcal{O}, \mathcal{A}, \pi_i^q$, $ \mathcal{M}, \mu$, $ \mathcal{G}_i)$, for
$~ i \in \{\text{Supervisor},\ \text{Journalist},\ \text{Legal},\ \text{Scientific}\}$. 
A separate \emph{answering agent} 
$A_{\text{Ans}} = (\mathcal{O}, \mathcal{A}_{\text{Ans}}, \pi^a, \mathcal{M}, \mu, \mathcal{G}_{\text{Ans}})$. 

\noindent
\textbf{Observation space:} 
  $\mathcal{O} = (\mathcal{N}_e, \mathcal{K}_{\text{LLM}}), \quad 
  \mathcal{N}_e = \{\mathcal{H}, \mathcal{T}_p, \mathcal{C}, \mathcal{I}_s, \mathcal{E}\}$,  
  where $\mathcal{H}$ = headline, $\mathcal{T}_p$ = preprocessed article text, $\mathcal{C}$ = extracted claim, $\mathcal{I}_s$ = image summary from image entities, $\mathcal{E}$ = retrieved evidence (credibility checks, reverse image search, KG facts), and $\mathcal{K}_{\text{LLM}}$ = LLM’s internal knowledge.
  
  \noindent \textbf{Action space:} We distinguish between different types of actions:
  $\mathcal{Q}: \text{space of probing questions}, ~\mathcal{A}_{\text{Ans}}: \text{space of concise answers}.
  $. Thus, the global action space is
  $\mathcal{A} = \mathcal{Q} \cup \mathcal{A}_{\text{Ans}}$. 
  
  \noindent \textbf{Policies:}  
  Each persona $A_i$ is guided by a \emph{questioning policy} 
  $\pi_i^q: \mathcal{O}\times\mathcal{M}\to \mathcal{Q}$, 
  which maps observations and memory to a probing question $q_i^t$. Agent $A_{\text{Ans}}$ uses an answering policy 
  $\pi^a: \mathcal{Q}\times\mathcal{O}\times\mathcal{M}\to \mathcal{A}_{\text{Ans}}$, 
  which generates concise, evidence-grounded answers $a_i^t$.
  
  \noindent
  \textbf{Memory} $\mathcal{M}$ is a structured contextual memory containing Q\&A history, abstracted insights, and summaries.
  
  \noindent 
  \textbf{Memory update function} $\mu: \mathcal{M} \times (\mathcal{Q}, \mathcal{A}_{\text{Ans}}) \to \mathcal{M}$ updates memory by synthesizing each new Q\&A pair.
  
  \noindent \textbf{Goals:} Each agent has a role-specific goal:  
  \begin{itemize}[left=0pt]
    \item $\mathcal{G}_{\text{Supervisor}}$: determine overall claim veracity,  
    \item $\mathcal{G}_{\text{Journalist}}$: probe credibility, bias, and contextual consistency,  
    \item $\mathcal{G}_{\text{Legal}}$: check regulatory or legal compliance,  
    \item $\mathcal{G}_{\text{Scientific}}$: validate factual and scientific correctness,  
    \item $\mathcal{G}_{\text{Ans}}$: generate concise, evidence-based answers to persona queries.  
  \end{itemize}

\noindent \textbf{Iterative reasoning}: The Q\&A cycle iterates over personas, incrementally enriching $\mathcal{M}$. At each round $t=1,\dots,\tau$, with observation {$o^t\sim\mathcal{O}$:} 
\begin{equation} \scriptsize
    \begin{split}
        q_i^t = \pi_i^q (o^t, M^{t-1}) \in \mathcal{Q}~; \qquad a_i^t = \pi^a(q_i^t, o^t, M^{t-1}) \in \mathcal{A}_{\text{Ans}}~; \\
        M^t = \mu(M^{t-1}, (q_i^t, a_i^t)).
    \end{split}
\end{equation}

\end{definition}

Beyond supporting classification, the shared memory $\mathcal{M}$ regulates the Q\&A process by preventing redundant questions, mitigating hallucinations, and enforcing insight diversity across personas. This ensures that MMAF systematically covers complementary verification dimensions, including factual validity, source credibility, political/regional bias, and legal compliance. The pseudo-code for the MMAF reasoning process is provided in Algorithm~\ref{algo:role_reasoning}.

\begin{algorithm}[!b]
\caption{Multi-step Multi-persona Agentic Reasoning}
\label{algo:role_reasoning}\small
\begin{algorithmic}[1]
\Require Number of rounds $\tau$; persona set $\mathcal{P}=\{\text{Supervisor},\text{Journalist},\text{Legal},\text{Scientific}\}$;~~
Observation components $\mathcal{N}_e=\{\mathcal{H},\mathcal{T}_p,\mathcal{C},\mathcal{I}_s,\mathcal{E}\}$ and internal knowledge $\mathcal{K}_{\text{LLM}}$
\Ensure Structured contextual memory $\mathcal{M}^{\tau}$
\State Define observation space $\mathcal{O} \gets (\mathcal{N}_e,\mathcal{K}_{\text{LLM}})$
\State Initialize memory $\mathcal{M}^{0} \gets \textsc{Init}(\mathcal{N}_e)$
\State Given persona question policies $\{\pi_i^q:\mathcal{O}\times\mathcal{M} \to \mathcal{Q}\}_{i\in\mathcal{P}}$ and answering policy $\pi^a:\mathcal{Q}\times\mathcal{O}\times\mathcal{M} \to \mathcal{A}_{\text{Ans}}$
\For{$t=1$ to $\tau$}
  \State Sample/compose observation $o^t \sim \mathcal{O}$
  \For{each $i \in \mathcal{P}$}
    \State \textbf{Question:} $q_i^t \gets \pi_i^q(o^t,\mathcal{M}^{t-1})$
    \State \textbf{Answer:} $a_i^t \gets \pi^a(q_i^t, o^t, \mathcal{M}^{t-1})$
    \State \textbf{Update memory:} $\mathcal{M}^{t} \gets \mu(\mathcal{M}^{t-1}, (q_i^t,a_i^t))$
  \EndFor
\EndFor
\State \textbf{return} $\mathcal{M}^{\tau}$
\end{algorithmic}
\end{algorithm}

\subsection{LLM Pseudo-Labeling: Policy, Decision Rules, and Output}
We formulate pseudo-labeling as a disciplined, evidence-aware decision stage where an LLM, constrained by a policy, operates over the structured final memory built from textual, visual, and KG evidence. The objective is to produce \textit{REAL} / \textit{FAKE} / \textit{UNCERTAIN} labels with calibrated confidence and a concise, auditable justification. To mitigate hallucinations and source blindness, the policy requires using the entire memory as the primary factual base, consulting general world knowledge only when necessary, and invoking optional tools (e.g., persuasion-based refinement) only under uncertainty. Evidence items are tagged as supporting or contradicting the claim and weighted by credibility and temporal relevance; the LLM aggregates these signals to choose a verdict and articulate a brief rationale that cites the memory. The output is a strict JSON record (verdict, confidence, justification), enabling downstream automation and supervision of a compact SLM while preserving transparency.

We obtain evidence-grounded pseudo-labels by querying an LLM classifier under a constrained policy: 
\begin{equation} \small 
(\mathcal{Y}_p, \mathcal{C}_s, \mathcal{J}) = \Phi~\bigl(\pi_c; \mathcal{M}^{\tau},
{\text{FewShots}}\bigr)
\end{equation}
where, $\mathcal{Y}_p \in \{\textit{REAL},\textit{FAKE},\textit{UNCERTAIN}\}$, $\mathcal{C}_s\in[0,1]$ is a confidence score, and $\mathcal{J}$ is a concise justification that cites items in $\mathcal{M}^{\tau}$ {(primary source)} for interoperability. {FewShots} guidance improves decision consistency; memory citation ensures auditability.
 
To handle cases where aggregated evidence is inconclusive, we apply a \emph{persuasion-based refinement} module only to {\textit{UNCERTAIN}} instances. We first detail the persuasion-based refinement, then return to the LLM policy decision rules.

\subsubsection{Persuasion-based Refinement}
We formalize persuasion-based refinement as a gated, evidence-augmenting procedure applied only to items initially labeled \textit{uncertain} by the LLM pseudo-labeler.

\noindent 
\emph{Gating:}
Let the uncertain set be defined as: 
\begin{equation} \scriptsize
\mathcal{U}=\{N \mid \mathcal{Y}_p(N) = \textit{UNCERTAIN} = \Phi~ \bigl(~\pi_c;  \mathcal{M}^{\tau},  \text{FewShots}~\bigr)\}.
\end{equation} 
This module is activated only when the LLM classifier cannot reach a stable verdict, minimizing unnecessary procedure. 
We begin by defining the {Persuasion Knowledge Model (PKM)} agent.

\begin{definition}[PKM Agent]\label{def:pkm_agent}
A PKM agent is an auxiliary LLM-based agent that detects and summarizes persuasion techniques to refine \textit{uncertain} pseudo-labels. It is defined as:
$A_{\text{PKM}} = \bigl(\mathcal{O},  \mathcal{A}_{\text{PKM}}$, $\pi_{\text{PKM}},  \mathcal{M},  \mu,  \mathcal{G}_{\text{PKM}}\bigr).$

\noindent\textbf{Observation space:}
$
\mathcal{O} = (\mathcal{N}_e,  \mathcal{K}_{\text{LLM}}),~~ 
\mathcal{N}_e=\{\mathcal{H},\mathcal{T}_p,\mathcal{C},\mathcal{I}_s,\mathcal{E}\}.
$ 
PKM primarily consumes $\mathcal{N}_e^{\text{PKM}}=\{\mathcal{C},\mathcal{T}_p\}$ together with the current memory ${\mathcal{M}}^\tau$ (which may already encode textual, visual, KG evidence and MMAF knowledge).

\noindent\textbf{Action space:}
$\mathcal{A}_{\text{PKM}}=\Bigl\{  z\in\{0,1\}^{T},~\ \mathcal{S}=\{s_t\},~\ u\in\mathds{N}^{G},~\ \Pi\in[0,1],~\ \mathcal{J}_{\text{PKM}} \Bigr\}$.
{Here, $T=23$ persuasion techniques (JRC taxonomy \cite{piskorski2023news}) and $G=6$ categories (\emph{justification, simplification, distraction, call, reputation attack}, and
manipulative wording}); $z$ is the technique-activation vector, $\mathcal{S}$ is extracted evidence spans (phrases), $u$ is category counts, $\Pi$ is a normalized persuasion index, and $\mathcal{J}_{\text{PKM}}$ is a concise narrative summary.

\noindent\textbf{Policy:}
$\pi_{\text{PKM}}:\ \mathcal{O}\times \mathcal{M}\ \longrightarrow\ \mathcal{A}_{\text{PKM}}$. 
Given $(\mathcal{C},\mathcal{T}_p,\mathcal{I}_s,\mathcal{M})$, the policy produces
(i) {$z,\mathcal{S}$} via detection over content; 
(ii) $u=\mathscr{A} z$ using a fixed category–technique incidence matrix ${\mathscr{A}}\in\{0,1\}^{G\times T}$;
(iii) $\Pi=\frac{\beta^\top u}{||\beta||_1}$ with $\beta \in \mathds{R}_+^G$ denoting the importance of the category;
(iv) $\mathcal{J}_{\text{PKM}}=\mathrm{Summarize}(z,\mathcal{S})$.
The PKM agent is \emph{gated}, it is invoked only if ${g_{\text{PKM}}}  = {\mathds{1}}[\mathcal{Y}_p=\textit{UNCERTAIN}]$; ${\mathds{1}}[.]$ is the indicator function.

\noindent\textbf{Memory:}
$\mathcal{M}$ stores structured Q\&A history, abstracted insights, and evidence summaries. 

\noindent\textbf{Memory update:} The update synthesizes persuasion signals into memory for downstream classification:
$
\mathcal{M}^{+}=\mu~ \bigl(\mathcal{M},\mathcal{A}_{\text{PKM}}\bigr).
$

\noindent\textbf{Goal:} To detect and explain persuasion to refine uncertain decisions and improve interoperability $\mathcal{G}_{\text{PKM}}$.

\noindent\textbf{Activation and effect:}
When {$ g_{\text{PKM}}=1$}, PKM runs once ({i.e.,} no Q\&A loop), updates $\mathcal{M}\to\mathcal{M}^{+}$, after which persuasion-aware re-scoring or re-querying of the pseudo-labeler is performed.
\end{definition}


\subsubsection{LLM Pseudo-Labeling Policy}
We formalize pseudo-labeling as a constrained decision function operating on the final memory. Let $\mathcal{M}^{\tau}$ be the structured memory, and let $\mathcal{E}^{+}$/$\mathcal{E}^{-}$ denotes supporting/contradicting evidence items, each endowed with a reliability score ${\mathcal{R}}(d_i)$ for $d_i \in {\mathcal{E}}_T \cup {\mathcal{E}}_I$. 

\emph{Policy constraints:}
We define a two-stage decision with strict source and tool gating: 
{\scriptsize{$$ 
{\Phi}:\ \mathcal{M}^{\tau}\ \mapsto\ ({\mathcal{Y}}_p,{\mathcal{C}}_s,{\mathcal{J}})~;~~ 
\tilde{\Phi}:\ \mathcal{M}^{+} \mapsto\ (\mathcal{Y}_p,\mathcal{C}_s,\mathcal{J}), \text{ if } g_{\text{PKM}} = 1
$$}}

\noindent
where, the \emph{preliminary} decision ${\Phi}$ {must} use only $\mathcal{M}^{\tau}$. The \emph{refined} decision $\tilde{\Phi}$ may additionally consult persuasion-based refinement, for uncertain cases.
The rationale is that hard gating reduces hallucinations and tool-induced drift by prioritizing curated memory as the primary knowledge source, while auxiliary tools are activated only when principled uncertainty arises.

\emph{Evidence-weighted aggregation and confidence:}
We aggregate reliability-weighted support/contradiction as: 
\begin{equation} \scriptsize
S^{+}=\sum_{d_i\in\mathcal{E}^{+}} \mathcal{R}(d_i)~;~~
S^{-}=\sum_{d_i\in\mathcal{E}^{-}} \mathcal{R}(d_i),   
\end{equation} 
weighting by source reliability yields robust signal integration.

\emph{Preliminary decision rule $\tilde{\mathcal{Y}}_p$:}
Let $\mathcal{E}^{\pm}_{H}$ be the subsets of high-reliable evidence (i.e., reliability score is more than a threshold {$\alpha_h$}) within $\mathcal{E}^{\pm}$; let $\gamma,\eta>0$ be hyperparameters.
\begin{equation}  \scriptsize 
\tilde{\mathcal{Y}}_p =
\begin{cases} 
\textit{REAL}, & \left(|\mathcal{E}^{+}_{H}|>0 \wedge |\mathcal{E}^{-}_{H}|=0\right)\ \ \text{or}\\[1pt]
    & \left(|\mathcal{E}^{+}_{H}|+|\mathcal{E}^{-}_{H}|=0 \wedge S^{+}_{\text{Med}}-S^{-}_{\text{Med}}\ge \gamma\right);\\[4pt]
\textit{FAKE}, & \left(|\mathcal{E}^{-}_{H}|>0 \wedge |\mathcal{E}^{+}_{H}|=0\right)\ \ \text{or}\\[1pt]
    &\left(\text{only low-reliable support}\right)\ \ \text{or}\ \ \left(\psi_{\text{impl}}(N)=1\right);\\[4pt]
\textit{UNCERTAIN}, & \left(|\mathcal{E}^{+}_{H}|>0 \wedge |\mathcal{E}^{-}_{H}|>0\right)\ \ \text{or}\ \ \left(S^{+}+S^{-}<\eta\right)
\end{cases}
\end{equation}
where, $S^{\pm}_{\text{Med}}$ {denote} the {medium}-reliable components (i.e., reliability score is between $\alpha_h$ and $\alpha_m$) of $S^{\pm}$, and $\psi_{\text{impl}}(N)\in\{0,1\}$ flags implausible/sensational content by deterministic checks.
High-reliable precedence encodes epistemic dominance; medium-only consistency covers cases with no high-reliable sources; uncertainty guards against conflicted or sparse evidence.

The module outputs a structured triplet in the form (\texttt{verdict}, \texttt{confidence}, \texttt{justification});  
$\texttt{verdict}: \mathrm{map}(\mathcal{Y}_p)\in\{\textit{REAL},\textit{FAKE}$, $\textit{UNCERTAIN}\}$, 
$\texttt{confidence}=\mathrm{round} \left(100 \mathcal{C}_s\right)$, $\texttt{justification}=\mathcal{J}$,  
with $\mathcal{J}$ restricted to cite items from $\mathcal{M}^{\tau}$ (and explicitly mark auxiliary knowledge, if any).
Deterministic rendering ensures auditability and supports SLM training.

\subsection{SLM Classifier for Final Output}
The LLM-based pseudo-labeling stage provides evidence-grounded reasoning and interpretability, but deploying LLMs alone is prone to overfitting on superficial linguistic cues. To balance interpretability with efficiency, we employ an SLM classifier that consumes both raw news content and the reasoning artifacts generated by the LLM, thus integrating factual content with structured justifications.  
Formally, let the structured input sequence be:   
\begin{equation} \scriptsize
x = \big[ \mathcal{H}   \oplus   \langle \texttt{SEP} \rangle   \mathcal{T}_p   \oplus   \langle \texttt{SEP} \rangle   \mathcal{I}_s   \oplus   \langle \texttt{SEP} \rangle   \mathcal{J} \big],
\end{equation}
where, $\mathcal{H}$ is the headline, $\mathcal{T}_p$ is preprocessed article text, $\mathcal{I}_s$ is the image summary, $\mathcal{J}$ is the justification from the LLM classifier, and $\oplus$ denotes concatenation with structural separators $\langle \texttt{SEP} \rangle$. This ensures modality-aware encoding and preserves logical boundaries between different information sources.  

We instantiate the classifier as a pre-trained transformer encoder $f_\theta$ (\textit{DistilRoBERTa-base}) parameterized by $\theta$. The classifier outputs the final decision label:  
$
    \mathcal{Y}_s = f_\theta(x) \in \{\textit{REAL}, \textit{FAKE}\}.
$

The parameters $\theta$ are optimized using standard cross-entropy loss.  
This design ensures 
(i) {efficiency}, since $f_\theta$ is lightweight compared to LLMs, 
(ii) {robustness}, as the SLM benefits from the distilled reasoning $\mathcal{J}$ rather than overfitting to raw text alone, and 
(iii) {generalization}, as the combined representation $(\mathcal{H},\mathcal{T}_p,\mathcal{I}_s,\mathcal{J})$ anchors decisions in both factual content and structured evidence.

\subsection{Model Deployment}
In inference, each news item undergoes preprocessing to yield structured content and evidence. These inputs are evaluated by a multi-persona reasoning module, which builds a consolidated memory passed to an LLM classifier. The LLM outputs preliminary labels with justifications, ensuring interpretability. For uncertain cases, a persuasion-based refinement module identifies manipulative cues to reduce ambiguity. Finally, the enriched representation, combining raw content and reasoning outputs, is processed by an SLM classifier, which produces the final decision. This modular design balances evidence-grounded reasoning (via LLM), disambiguation (via persuasion refinement), and computational efficiency with generalization (via SLM).

\section{Experiments}\label{sec:result}

All experiments were conducted on a system equipped with PyTorch 2.5.1, CUDA 12.1, NVIDIA A100-SXM4-40GB GPU, AMD EPYC 7742 64-core CPU with 1 TB RAM. We used the Google Cloud Custom Search API to retrieve external evidence, and the OpenAI GPT-4o mini for LLM-based inference. 
For the {final verdict}, DistilRoBERTa-base \cite{distilroberta-base} was finetuned as SLM to obtain \textit{REAL} or \textit{FAKE}.

\subsection{Datasets and Evaluation Metrics}

\subsubsection{Datasets} 
To evaluate our model, we used three widely adopted fake news datasets: PolitiFact \cite{shu2020fakenewsnet}, GossipCop \cite{shu2020fakenewsnet}, and MMCoVaR \cite{chen2021mmcovar}. PolitiFact contains fact-checked political claims with veracity labels, while GossipCop consists of celebrity and entertainment news articles annotated as \textit{REAL} or \textit{FAKE}. MMCoVaR is a multimodal dataset focused on COVID-19 vaccine-related misinformation, comprising textual posts along with associated images. We adopted a $7{:}2{:}1$ train–validation–test split, and all reported results are based on the test set.


\subsubsection{Evaluation Metrics} 
To evaluate the effectiveness of the model, we employ four widely adopted metrics: Accuracy \% ($Acc$), F1 Score \% ($F1$), Precision \% ($Pre$), and Recall \% ($Rec$). In addition, we report an improvement metric that measures the percentage gain in our method over the second-best baseline. 
To evaluate the quality of the generated rationale (LLM justification $\mathcal{J}$), we employ three complementary metrics: 
{(i)} the {Utility} metric $\mathcal{U}$, capturing the effectiveness of the rationale as the performance difference between fine-tuned SLMs trained with and without LLM justification $\mathcal{J}$; 
{(ii)} the 
{Leakage-Adjusted Simulatability ($LAS$)} metric~\cite{las_metric}, 
assessing {faithfulness}, how well $\mathcal{J}$ aids in simulating the verdict without leaking label information, computed as
$\mathds{1}[\mathcal{Y}_s \mid \mathcal{H}, \mathcal{T}_p, \mathcal{I}_s, \mathcal{J}] - \mathds{1}[\mathcal{Y}_s \mid \mathcal{H}, \mathcal{T}_p, \mathcal{I}_s]$, where $\mathds{1}[\cdot]$ is an indicator function that equals $1$ if the model correctly predicts the true label and $0$ otherwise; 
and 
{(iii)} the {Rationale Evaluation via conditional $\mathcal{V}$-information ($REV$)} metric~\cite{rev_metric}, quantifying \textit{informativeness}, the additional label-relevant information provided by $\mathcal{J}$, computed as the log-likelihood difference of the true label $\mathcal{Y}$ between two models $f_\theta$ and $f'_\theta$ trained with $\mathcal{J}$ and without $\mathcal{J}$ as inputs, respectively:
$\log f_{\theta}([\mathcal{H}, \mathcal{T}_p, \mathcal{I}_s], \mathcal{J})(\mathcal{Y}) -\log f'_{\theta}([\mathcal{H}, \mathcal{T}_p, \mathcal{I}_s])(\mathcal{Y})$. 
\subsection{Results and Analysis} 
To benchmark our approach, AMPEND-LS, we compare it against two sets of baselines, 
\emph{(i)} LLM-based methods: LMDD \cite{Kheddache2025} and FactAgent \cite{Li2024}, 
and 
\emph{(ii)} Learning-based methods, including five recent state-of-the-art (SOTA) methods, 
MFCL \cite{CHEN2025112800}, 
BREAK \cite{yin2025break}, 
MGCA \cite{Guo_Ma_Zeng_Luo_Zeng_Tang_Zhao_2025},
EARAM \cite{10.1145/3696410.3714532}, and 
FakingRecipe \cite{10.1145/3664647.3680663}.

As shown in Table~\ref{tab:sota}, AMPEND-LS consistently outperforms all baselines across the three datasets—PolitiFact, GossipCop, and MMCoVaR—achieving the highest accuracy, $F1$, precision, and recall in each case. The most substantial gains are observed on GossipCop, with improvements of +6.17\% in accuracy and +9.07\% in $F1$ over the strongest baseline, demonstrating the framework’s effectiveness in handling entertainment-oriented and rumor-driven content. On PolitiFact and MMCoVaR, steady improvements further confirm strong generalization across both text-centric and multimodal misinformation settings. 
Unlike several competing methods that exhibit imbalanced precision–recall trade-offs (e.g., higher precision but lower recall in MGCA, or vice versa in MFCL), AMPEND-LS maintains consistently aligned performance across all evaluation metrics. This balanced behavior can be attributed to the complementary integration of agentic multi-persona reasoning, persuasion-aware refinement, and credibility-guided multimodal evidence fusion.  
While certain baselines such as BREAK perform competitively in specific domains (e.g., MMCoVaR), their performance does not generalize consistently across datasets. In contrast, AMPEND-LS demonstrates stable and uniform gains across heterogeneous domains, suggesting improved robustness under distributional variation. 
The close alignment among accuracy, precision, recall, and $F1$ further indicates stable decision boundaries and reduced susceptibility to overfitting. Overall, these results establish AMPEND-LS as a robust, adaptable, and evidence-grounded framework capable of effectively addressing diverse misinformation scenarios with enhanced transparency and reliability.

\begin{table}[!t]
    \centering
    \scriptsize 
    \caption{Comparison with SOTA methods}
    \begin{adjustbox}{width=\linewidth}
    \begin{tabular}{ll cccc} \hline
        & Method & $Acc$ & $F1$ & $Pre$ & $Rec$ \\ \hline \hline 
    \multirow{9}{*}{\rotatebox[origin=c]{90}{PolitiFact}} 
        & LMDD \cite{Kheddache2025} & 	80.86 & 	75.99 & 	80.05	& 72.34 \\
    	& MFCL \cite{CHEN2025112800} 	& 81.82	&  82.52	& 81.61	&  83.47 \\
    	& BREAK \cite{yin2025break} &  82.14	& 82.03	& 81.79	& 82.29 \\
    	& MGCA \cite{Guo_Ma_Zeng_Luo_Zeng_Tang_Zhao_2025} 	&  82.14	& 81.10	&  82.14 & 80.09 \\
    	& EARAM \cite{10.1145/3696410.3714532} & 	82.93	& 81.11	& 80.90	& 81.32 \\
    	& FakingRecipe \cite{10.1145/3664647.3680663} & 83.33	& 80.07	& 80.83	& 79.44 \\ 
        & FactAgent\cite{Li2024} 
        & \underline{88.12} & \underline{88.51} & \underline{87.52} & \underline{88.11} \\  [0.07cm]
        \cline{2-6}
    	& \textbf{AMPEND-LS (Our)} & \textbf{92.18 ± 1.93}	& \textbf{91.64 ± 2.12} & \textbf{92.06 ± 1.85} & \textbf{91.40 ± 2.43} \\ 
    	& Improvement (\%)	& 4.61	& 3.53 & 5.19 & 3.73 \\ \hline
        
    \multirow{9}{*}{\rotatebox[origin=c]{90}{GossipCop}} 
        & LMDD \cite{Kheddache2025} & 65.17	& 71.05 & 78.21 & 	65.10 \\
    	& FakingRecipe \cite{10.1145/3664647.3680663} 	& 80.26	& 80.27	& 80.21	& 80.26 \\
    	& MFCL \cite{CHEN2025112800}	& 81.81	& 81.35	& 81.65	& 81.06 \\
    	& BREAK \cite{yin2025break} & 82.00	& 82.11	& 82.51	& 81.73 \\
        & FactAgent \cite{Li2024} 
        & 83.00 & 83.00 & 83.04 & 83.02 \\
    	& EARAM \cite{10.1145/3696410.3714532} & 85.00	&  \underline{85.49}	& 86.00 &  \underline{85.00} \\
    	& MGCA \cite{Guo_Ma_Zeng_Luo_Zeng_Tang_Zhao_2025} 	&  \underline{88.00}	& 84.09	&  \underline{88.31}	& 80.26 \\ [0.07cm]
        \cline{2-6}
        
    	& \textbf{AMPEND-LS (Our)} & \textbf{93.43 ± 1.53} & \textbf{93.24 ± 1.59}& \textbf{93.44 ± 1.48}	& \textbf{93.29 ± 1.55} \\ 
        
    	& Improvement (\%)	& 6.17 & 9.07 & 5.81 & 9.75 \\ \hline

    \multirow{9}{*}{\rotatebox[origin=c]{90}{MMCoVaR}} & 
         LMDD \cite{Kheddache2025} & 	64.62 & 	63.54& 	78.61& 	53.40 \\
    	& FakingRecipe \cite{10.1145/3664647.3680663} &  79.96	& 78.61	& 79.35 & 77.92 \\
    	& MFCL \cite{CHEN2025112800}	& 82.06	& 78.19	& 77.82	& 78.57 \\
        & FactAgent\cite{Li2024} 
        & 86.97  & 85.97 & 86.33 & 86.97 \\
    	& MGCA \cite{Guo_Ma_Zeng_Luo_Zeng_Tang_Zhao_2025} 	& 90.00	& 88.34	& 88.19 & 	88.48 \\
    	& EARAM \cite{10.1145/3696410.3714532} &  90.36	& 88.51	& 88.32 & 88.70 \\
    	& BREAK \cite{yin2025break} &  \underline{91.89}	&  \underline{89.27}	&  \underline{88.73} &  \underline{89.83} \\ 
        [0.07cm] 
        \cline{2-6} 
    	& \textbf{AMPEND-LS (Our)}	& \textbf{96.46 ± 1.11}	& \textbf{95.56 ± 1.31} & \textbf{96.30 ± 1.48} & \textbf{94.96 ± 1.55} \\ 
        & Improvement (\%)& 	4.97	& 	7.05 & 8.53 & 5.71 \\ \hline
\multicolumn{6}{r}{Improvement (\%) of \textbf{AMPEND-LS} over the \underline{second-best} method}
    \end{tabular}
     \end{adjustbox}
    \label{tab:sota}
\end{table}


\subsection{Ablation Study}

To evaluate the contribution of each component in AMPEND-LS, we conduct a systematic ablation study on PolitiFact, GossipCop, and MMCoVaR. The results are summarized in Table~\ref{tab:module_ablation}. The key observations are as follows:

\emph{(i) Impact of Persuasion Analysis (PKM):}  
The PKM module alone yields relatively modest performance (e.g., $Acc:$  73.13 on PolitiFact, 67.90 on GossipCop, and 71.94 on MMCoVaR), indicating that persuasion signals by themselves are insufficient for reliable misinformation classification. However, removing PKM from the full pipeline (w/o PKM) results in consistent performance degradation compared to AMPEND-LS (e.g., –6.41 points on PolitiFact, –7.15 on GossipCop, and –10.49 on MMCoVaR in accuracy). This demonstrates that PKM plays a complementary role in resolving uncertain or rhetorically manipulative cases, strengthening final decisions when integrated with evidence and reasoning modules.

\emph{(ii) Impact of the Evidence Module:}  
The Evidence-only configuration establishes a meaningful baseline (e.g., $Acc:$  74.26 on PolitiFact, 79.67 on GossipCop, and 80.04 on MMCoVaR), confirming that credibility-aware retrieval provides substantial factual grounding. However, it remains significantly below the integrated AMPEND and AMPEND-LS models. For instance, on PolitiFact, $F1$ improves from 73.16 (Evidence) to 88.24 (AMPEND) and further to 91.64 (AMPEND-LS). This highlights that reliability scoring, temporal filtering, and multimodal grounding are necessary but not sufficient without structured reasoning and refinement.

\emph{(iii) Impact of Agentic Reasoning:}  
The standalone Agentic module outperforms both PKM and Evidence alone (e.g., $Acc:$  79.92 on PolitiFact, 83.14 on GossipCop, 81.20 on MMCoVaR), demonstrating that structured multi-step reasoning significantly enhances claim analysis. Nevertheless, it still underperforms compared to AMPEND and AMPEND-LS, indicating that reasoning benefits substantially from explicit credibility signals and persuasion-aware refinement. Thus, reasoning alone improves structure, but integration ensures robustness.

\emph{(iv) Single-Persona vs. Multi-Persona (ASPEND vs. AMPEND):}  
The single-persona model (ASPEND) achieves strong performance (e.g., $Acc:$  84.13 on PolitiFact, 88.80 on GossipCop, 86.28 on MMCoVaR), yet consistently trails the multi-persona AMPEND (88.54, 92.68, and 91.27 respectively). This confirms that modeling complementary social roles (journalistic, legal, scientific, supervisory) enhances perspective diversity, reduces bias, and improves decision reliability.

\emph{(v) AMPEND vs. AMPEND-LS (Effect of SLM Distillation):}  
Integrating SLM distillation yields further consistent gains across all datasets. For example, on MMCoVaR, accuracy increases from 91.27 (AMPEND) to 96.46 (AMPEND-LS), with corresponding improvements in $F1$, precision, and recall. Similar gains are observed on PolitiFact and GossipCop. These results indicate that the SLM serves as a stabilizing decision layer, mitigating LLM uncertainty while preserving evidence-grounded reasoning. The synergy between agentic reasoning and lightweight classification enhances both performance and deployment scalability.

Overall, the ablation study confirms that each component (credibility-aware evidence retrieval, multi-persona reasoning, persuasion-aware refinement, and SLM distillation) contributes complementary benefits. Their integration yields a system that is not only more accurate but also more stable and robust across heterogeneous misinformation domains.

\begin{table}[!t]
    \centering
    \scriptsize
    \caption{Module-wise ablation study}
    \begin{tabular}{cl cccc} \hline
       Dataset  & Method & $Acc$ & $F1$ & $Pre$ & $Rec$ \\ \hline \hline 
        \multirow{7}{*}{\rotatebox[origin=c]{90}{PolitiFact}}  
          & 	PKM & 	73.13 & 	73.49 & 	81.23 & 	67.10 \\
        	 & Evidence & 	74.26 & 	73.16 & 	79.12 & 	68.05 \\
        	 & Agentic & 	79.92 & 	78.16 & 	81.40 & 	75.17 \\
        	 & w/o PKM & 	85.77 & 	86.28 & 	87.31 & 	85.28 \\
        	 & ASPEND & 	84.13 & 	84.18 & 	87.32 & 	81.26 \\
        	 & AMPEND & 	88.54 & 	88.24 & 	90.26 & 	86.32 \\
        	 & \textbf{AMPEND-LS} & 	\textbf{92.18} & 	\textbf{91.64} & 	\textbf{92.06} & \textbf{91.40}  \\ \hline
             
        \multirow{7}{*}{\rotatebox[origin=c]{90}{GossipCop}} 
          & PKM & 	67.90 & 	66.67 & 	72.02 & 	62.07 \\
        	 & Evidence & 	79.67 & 	79.71 & 	80.06 & 	79.37 \\
        	 & Agentic & 	83.14 & 	83.65 & 	84.20 & 	83.11 \\
        	 & w/o PKM & 	86.28 & 	86.03 & 	86.13 & 	85.98 \\
        	 & ASPEND & 	88.80 & 	89.01 & 	89.08 & 	88.92 \\
        	 & AMPEND & 	92.68 & 	92.56 & 	93.04 & 	92.10 \\
        	 & \textbf{AMPEND-LS} & 	\textbf{93.43} & 	\textbf{93.24} & 	\textbf{93.44} & \textbf{93.29} \\ \hline

        \multirow{7}{*}{\rotatebox[origin=c]{90}{MMCoVaR}} 
          & PKM &  71.94	& 63.10	& 76.21 & 72.20  \\
        	 & Evidence & 80.04 & 74.10	& 78.12	& 80.02	 \\
        	 & Agentic & 81.20	& 76.01	& 78.15 & 81.10	 \\
        	 & w/o PKM & 85.97 	& 85.02	& 86.20	& 86.01	 \\
        	 & ASPEND &  86.28	& 85.12	& 85.01	& 86.20	 \\
        	 & AMPEND &  91.27	& 89.19	& 92.11	& 86.45	 \\
        	 & \textbf{AMPEND-LS} & \textbf{96.46}	& \textbf{95.56}	& \textbf{96.30}	& \textbf{94.96}	 \\ \hline
             
            \end{tabular}
    \label{tab:module_ablation}
\end{table}

\begin{figure}[!b]
        \centering
        \begin{tikzpicture} 
            \begin{axis}[
                width=0.4\textwidth,
                height=0.2\textwidth,
                ymin=85, ymax=91,
                xtick={1,2,3,4},
                xticklabels={1,2,3,4},
                ytick={85,87,89,91},
                xlabel={Rounds},
                ylabel={Performance (\%)},
                legend style={at={(0.78,0.40)}, anchor=north, legend columns=2, font=\scriptsize, draw=none, fill=none},
                legend image post style={scale=0.8},
                line width=1pt,
                mark size=3pt,
                ymajorgrids=true,
                xmajorgrids=true,
                grid style=dashed,
                tick label style={font=\Large}
            ]

            \addplot[color=blue, mark=*] coordinates 
            {(1,88.54) (2,88.56) (3,88.93) (4,89.30)};
            \addlegendentry{$Acc$}

            \addplot[color=orange, mark=square*] coordinates 
            {(1,87.34) (2,87.32) (3,88.24) (4,88.54)};
            \addlegendentry{$F1$}

            \addplot[color=red, mark=triangle*] coordinates 
            {(1,90.26) (2,90.21) (3,89.41) (4,90.03)};
            \addlegendentry{$Pre$}

            \addplot[color=teal, mark=diamond*] coordinates 
            {(1,86.32) (2,86.05) (3,87.62) (4,87.82)};
            \addlegendentry{$Rec$}

            \end{axis}
        \end{tikzpicture}
        \caption{Performance across multi-rounds Q\&A on PoltiFact.}
        \label{fig:rounds}
\end{figure}
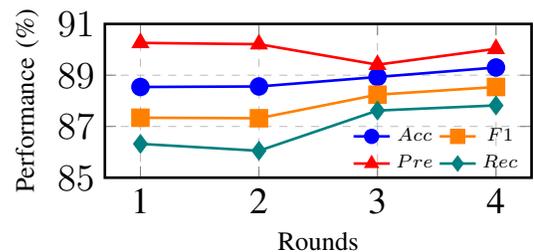

\subsection{Sensitivity to Reasoning Depth in Multi-Persona Framework}

Fig.~\ref{fig:rounds} illustrates the impact of increasing the number of multi-persona Q\&A rounds on the PolitiFact dataset. AMPEND-LS exhibits stable and incremental performance improvements as reasoning depth increases. Specifically, $Acc$ improves from 88.54 at round 1 to 89.30 at round 4 (+0.76 percentage points), $F1$ increases from 87.34 to 88.54 (+1.20 pp), and $Rec$ rises from 86.32 to 87.82 (+1.50 pp). Meanwhile, $Pre$ remains consistently high (approximately 90\%) with only minor fluctuations across rounds. These results indicate that additional reasoning cycles help refine evidence aggregation and improve the precision–recall balance. 
However, the marginal performance gains diminish beyond the third round. Since each reasoning cycle incurs additional LLM calls, increasing the number of rounds directly impacts latency and computational cost. From a systems perspective, 2–3 rounds provide a favorable trade-off, capturing most of the performance gains while avoiding unnecessary overhead. 
This controlled reasoning-depth analysis further supports the deployment design of AMPEND-LS, where adaptive round selection can dynamically balance accuracy, robustness, and computational efficiency in real-time misinformation monitoring scenarios. 
Although only PolitiFact is shown here for clarity, we conducted analogous experiments on GossipCop and MMCoVaR and observed consistent trends, confirming that the reasoning-depth trade-off generalizes across datasets.

\subsection{Information-Theoretic Evaluation of LLM-Justification}

To assess the effectiveness of LLM-generated justifications, we evaluate their contribution along three complementary dimensions: $REV$ \cite{rev_metric}, $LAS$ \cite{las_metric}, and the utility performance gains 
$\mathcal{U}_i$, where $i {\in} \{\text{$Acc$}, \text{$F1$}, \text{$Pre$}, \text{$Rec$}\}$, as reported in Table~\ref{tab:rev_results}. 
Across all datasets, AMPEND-LS achieves consistently positive $REV$ and $LAS$ scores, indicating that the generated justifications provide additional label-relevant information (higher $REV$) while maintaining faithful alignment with model predictions (positive $LAS$). Furthermore, all utility metrics are positive, demonstrating that incorporating LLM-generated rationales into SLM fine-tuning yields measurable improvements in predictive performance. 
Among the three datasets, PolitiFact exhibits the largest gains in both information-theoretic measures and downstream utility (e.g., $\mathcal{U}_{F1}=11.72$, $\mathcal{U}_{Rec}=13.62$), suggesting that justification signals are particularly beneficial in fact-centric political contexts. GossipCop and MMCoVaR also show consistent, though comparatively moderate, improvements, confirming that the benefits of justification integration generalize across entertainment and health misinformation domains. 
Overall, these findings show that faithful and informative LLM justifications contribute meaningful task-relevant signals that simultaneously improve downstream performance and interpretability, effectively bridging explainability and utility in aligned SLM models.

\begin{table}[!h]
    \centering
    \caption{LLM justification analysis}
    \scriptsize
    \begin{adjustbox}{width=0.43\textwidth}
        \begin{tabular}{lc c cccc}
            \hline 
            {Dataset} & {$REV$} & {$LAS$} &  {$\mathcal{U}_{\text{$Acc$}}$} & {$\mathcal{U}_{\text{$F1$}}$} & {$\mathcal{U}_{\text{$Pre$}}$} & $\mathcal{U}_{\text{$Rec$}}$ \\ \hline \hline 
            PolitiFact & 0.3885	& 0.0976 & 9.76	& 11.72	& 6.58	& 13.62 \\
            GossipCop  & 0.2460	& 0.0789 & 7.89	& 7.98	& 6.95	& 7.89 \\
            MMCoVaR    & 0.1845	& 0.0363 & 3.63	& 4.90	& 3.09	& 6.04 \\ \hline
        \end{tabular}
    \end{adjustbox}
    \label{tab:rev_results}
\end{table}

Appendix A outlines the configuration details of AMPEND-LS, Appendix B presents the qualitative analysis, and Appendix C provides the cross-domain analysis.

\section{Limitations and Future Work}
\label{limitation}
While AMPEND-LS achieves strong performance and interpretability across diverse benchmarks, several limitations point to promising directions for future research. The framework’s reliance on external APIs (e.g., web and reverse image search) may introduce latency, coverage gaps, or dependency risks. Future work may explore retrieval-agnostic or locally cached evidence pipelines to improve robustness and deployment reliability. 

The multi-persona reasoning mechanism, although enhancing accuracy and transparency, requires repeated LLM interactions that increase computational and monetary cost. Designing adaptive, cost-aware policies to dynamically control reasoning depth presents a promising avenue for scalable real-time deployment. 

Furthermore, our experiments are limited to three English-centric datasets (PolitiFact, GossipCop, and MMCoVaR). Extending evaluation to multilingual, low-resource, and culturally diverse misinformation scenarios is essential for broader applicability. Current credibility scoring relies on third-party databases such as MBFC, which may not fully capture the dynamics of evolving global media ecosystems; adaptive credibility estimation that continuously incorporates emerging sources could strengthen long-term resilience.

Finally, although persuasion detection and multimodal evidence fusion enhance robustness, adversarial manipulations—such as AI-generated deepfakes and coordinated misinformation campaigns—remain open challenges. Incorporating adversarial training, generative counterfactual reasoning, and stronger multimodal alignment mechanisms may further improve system resilience.

By acknowledging these limitations as natural extensions rather than shortcomings, we emphasize that AMPEND-LS provides a solid and adaptable foundation. Addressing these directions will enhance scalability, inclusivity, and robustness, strengthening the system’s role as a deployable and trustworthy solution for combating misinformation in dynamic real-world environments.

\section{Conclusion}\label{sec:conclusion}
This work introduced AMPEND-LS, an agentic multi-persona framework for evidence-aware fake news detection that integrates multimodal evidence retrieval, structured reasoning, and hybrid LLM–SLM classification within a unified architecture. By combining multi-perspective, role-aware analysis with lightweight SLM adaptation, the framework achieves state-of-the-art performance while maintaining transparency, robustness, and scalability.
Extensive experiments across political, entertainment, and health misinformation domains demonstrate consistent improvements in predictive accuracy and interpretability. The modular design further provides a strong foundation for extension to multilingual, low-resource, and adversarially robust settings, paving the way for trustworthy deployment in real-world, high-stakes information ecosystems.

\balance 

\bibliographystyle{IEEEtran}  
\bibliography{6_references}


\section*{Supplementary Appendix}
Appendices A, B, and C are provided in \url{https://github.com/csksuraj17/AMPEND-LS_Supp}

\end{document}